# Outlier Detection Techniques for SQL and ETL Tuning

Saptarsi Goswami
AKCSIT
Calcutta University,
Kolkata, India

Samiran Ghosh
AKCSIT
Calcutta University,
Kolkata, India

Amlan Chakrabarti
AKCSIT
Calcutta University,
Kolkata, India

## ABSTRACT

RDBMS is the heart for both OLTP and OLAP types of applications. For both types of applications thousands of queries expressed in terms of SQL are executed on daily basis. All the commercial DBMS engines capture various attributes in system tables about these executed queries. These queries need to conform to best practices and need to be tuned to ensure optimal performance. While we use checklists, often tools to enforce the same, a black box technique on the queries for profiling, outlier detection is not employed for a summary level understanding. This is the motivation of the paper, as this not only points out to inefficiencies built in the system, but also has the potential to point evolving best practices and inappropriate usage. Certainly this can reduce latency in information flow and optimal utilization of hardware and software capacity. In this paper we start with formulating the problem. We explore four outlier detection techniques. We apply these techniques over rich corpora of production queries and analyze the results. We also explore benefit of an ensemble approach. We conclude with future courses of action. The same philosophy we have used for optimization of extraction, transform, load (ETL) jobs in one of our previous work. We give a brief introduction of the same in section four.

## General Terms

Data Mining,

## Keywords

Outlier Detection, ETL Tuning, Query Tuning

## 1. INTRODUCTION

DBMS is center-piece for both Online Analytical (OLAP) as well as Online Transactional Processing system (OLTP). We would focus more an OLAP queries as the volume of the data and diversity of workload is much richer in a Data warehouse centric environment i.e. an OLAP system. Ref [1] observes a small sized data warehouse can have five terabytes (TB) of data. There are as many as sixteen major proven DBMS vendors (Oracle, Teradata, Microsoft, IBM, Asterdata, SAP/Sybase etc.). Datawarehouse has evolved from a traditional business intelligence platform to encompass Operational BI (Business Intelligence), operational analytics and performance management. There has been increased importance to near real time data. There have been emerging trends like emphasis on appliance based solution; column based stores and massively parallel architecture, in Memory etc. However, still increased demand for optimization techniques and performance enhancement remain top of primary forces to impact DW DBMS markets in 2011[1]. Another important force is ability to support mixed workloads. Both of this motivates outlier detection technique on the SQL queries for a better understanding as well as optimization.

There are very standard ways to optimize loading and retrieval of data. First part of it depends on database engine; as example, some database engines rely on hash based data distribution (Teradata) resulting in an "shared nothing" architecture, where a traditional DBMS engine will focus more on query rewrite, materialized views, various kind of partitioning and indexing strategy. A column based data base again seeks to leverage high rate of compression because of similar domain of values. The second part is underlying software and hardware (A RAID solution, multi-processor etc.). The third part depends on the database designers as they decide on the indexing and partitioning strategies. The fourth part is more dependent on the basic and advanced users. Their adherence to standard best practices has lot to do with optimal performance of a query.

The scope of the paper is certainly addressing the third and more specifically the fourth part of the optimization. There are plethora of checklists, tools, and review processes present to enforce the same. There are three issues with the same, one, often they are so subjective and commonsensical it is hard to get them validated. While validating against a rule like usage of functions in a where clause is trivial a best practice like using as many as temporary tables as possible for a large data processing job is impossible to validate. Two, because of the above reason it becomes manual, hence time consuming and error-prone. We refer the above way of validation as 'White Box Technique' simply because we need to inspect how it is done? Thirdly, standards are continuously evolving because of both a product version upgrade as well as increased understanding of best practices with time. As a result, there might be inefficient processes running in production which are inappropriately using both the hardware and software. Its effect can be manifested in two ways, one in increased information delivery latency, two in taking incorrect and untimely decision on capacity.

We seek to address this very area in this paper. All this DBMS engines stores execution related attributes in system tables. We take a black-box view and try to look at the queries as groups. We observe the general patterns, behaviors of the queries and look at the queries which are outlying from the crowd. We explore them further to identify corrective action. It can also provide guidance on the optimal scheduling of parallel/serial, dependent/independent queries by studying the temporal behavior. However we have not covered the same in the scope of the paper.

The organization of the paper is as follows: Section II briefs the related works in this area. Section III introduces basic concepts on Outlier Detection, and Section IV discusses on summary of our work in ETL optimization. Section V provides a brief overview on query optimization. Section VI details on the experiment setup and results and Section VII discusses on challenges, future course of work and conclusion.

## 2. RELATED WORK

There are two domains of work related to the paper. One, various query optimization techniques Two, different outlier detection techniques and their applications. [2], [3] refers in details of outlier detection taxonomy in terms of application domain, algorithms, input data, output, level of supervision, evaluation techniques etc. There has been research work addressing specific problems in the domain like high dimensionality applying principal component analysis (PCA) [4] or online detection from sensor data [5]. Application of outlier detection techniques are in fraud detection [6], network intrusion detection [7], medical health to name a few.

Extensive work in Query optimization has started form 1970. [8],[9] can be referred for a detailed understanding on the query optimization especially for a DBMS system. This is a well developed and researched area and now recent works are more specific like a 'Query Hint Framework' in [10] or in optimization of XML Queries [11].

While Query optimization takes place during and pre execution, result of the queries in terms of various cost components is not analyzed. We take this actual execution costs and use various outlier detection techniques to find out outlying queries. We have done similar outlier detection on ETL Execution traces in [12]. We give a brief overview on the same in section four.

## 3. OUTLIER

[13] Defines an outlying observation, or outlier, is one that appears to deviate considerably from other members of the sample in which it occurs. Another definition as observed in [14] is, it is an observation that deviates so much from other observations as to arouse suspicion that it was generated by a different mechanism. In Figure 1 we can perceive N1 and N2 are normal regions where as O1 is an outlier.

Outliers can be classified in two major ways one where a particular instance is an outlier. Alternatively a sequence of observations can also be an outlier. Further, for individual outlier an additional notion of context can be introduced by looking at spatial and temporal attributes. The input data is similar to any regular data mining task and can have binary, categorical (nominal and ordinal), discrete and continuous data types. The output of an outlier detection task would be either a level (Outlier or normal) or a score. The later is more preferable as it gives an idea on level of outlyingness. The methods can be supervised, semi-supervised and unsupervised. Problem with supervised and semi-supervised data is availability of labeled data for training and normal behavior can evolve over time. Unsupervised technique is better-off due to this, however, suffers from relatively higher false alarm rate.

There are different ways of detecting outlier namely classification based, Nearest Neighbor based, clustering based and statistics based etc.. Nearest Neighbor can be further classified in distance based and density based techniques. The statistics based techniques will be either parametric (assumes a data distribution model) or non-parametric. They are applied in various domains like fraud detection, intrusion detection, medical data, sports, novel topic detection etc. For a detailed overview on the outlier we can refer [2]. Outliers are also referred as anomaly, novelty, exception, surprise etc.

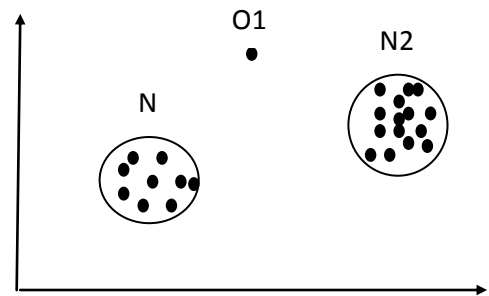

**Fig. 1 Outlier in 2D Space**

## 4. ETL OPTIMIZATION

ETL (Extract, Transform, and Load) layer is one of the most important layers in the Data warehousing (DW) Scheme of things. Companies spend billions of dollars in getting clean, unambiguous data in their data warehouse. [15] Observes 70% of the effort and time building a data warehouse goes into this extracting, cleaning, conforming, transforming and loading data. Basic philosophy of our work in [12] have been expressing ETL jobs as vectors in multi dimensional space and determine the jobs which are furthest from the group. There was a need of information extraction module as the ETL logs are text files with unstructured data. We followed the following steps:

1. We conducted a survey among a group of developers, architects to shortlist few priority parameters of an execution trace.

2. We construct information extract module for obtaining the parameters from the log files.

3. We apply clustering algorithm on a set of 500 + production logs.

4. We identify the ones in smaller clusters as outliers.

Below is the summary of the result

To summarize, our algorithm helped us to narrow down our investigation scope from 530 to 44, which is 8% of the overall. Between them 2 clusters with minimum population are the actual outliers because of the connection type or the inefficient source query, where as the other 3 are outliers because of huge number of rows than the rest.

Our approach of extracting metrics from ETL log was very simple yet immediate benefits can be achieved from the same. This does not really need any significant additional investment from the organization. All this information is already captured. Neither the text parsing application nor the data mining application involves noticeable cost. [16] Discusses on outlier detection for process logs, however the focus is on finding the structural pattern, the relationship among the activities and then finding outlier. Output of the clustering algorithm is shown below in Fig 2.

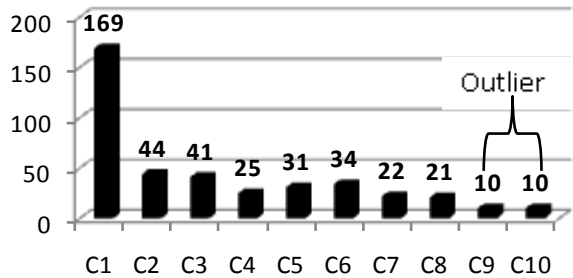

**Fig. 2: Cluster of ETL jobs**

In our current work we look at SQL Queries. We normalize the data values and we propose to use an ensemble method rather than a single method.

## 5. QUERY OPTIMIZATION

In this section we take a look at various ways of obtaining optimal performance from a query. While much of it depends on the optimizer and underlying hardware and software, a significant part can be attributed to a proper database design and correctly formulated queries. Query optimization is one of the major steps in query processing. For a particular query there can be variety of methods to get the results. Each method will have different query cost. The task of optimization is picking up the one with lowest cost. As there can be numerous numbers of ways to obtain the result, heuristics is employed for pruning. The cost can be broadly classified in the following five areas. A) Access cost to Secondary Storage B) Storage cost (for intermediate files) C) CPU or computation cost D) Memory usage E) Communication cost. From a practical point of view the major emphasis is on minimizing the access cost to secondary storage. The costs are calculated on the basis of statistics on relations and indexes (Number of data pages in a relation, number of data pages in an index), selectivity of predicates etc. Because of inaccurate information an inappropriate plan might get selected. To avoid the same, query hints can be used. This part is mostly taken care of by optimizer, and most of the commercial optimizers are matured enough to handle different kind of work loads. Another part of query optimization is dependent on the designer decision in the way he/she selects indexing, partitioning strategy etc. Certainly the query performance can be augmented by using more memory or processor. The next part is conformance to best practices by the individual developer.

Now this is the part we see scope of inefficiencies as often standards are evolving, as well as the validation part is mostly manual hence error prone and time consuming and often the existing queries are resource heavy which never comes under the scanner. We address this particular area in our paper, where we examine the execution characteristics of the queries and try and find the anomalous or the outlying queries. The reason might be varied like, incorrect query formulation, improper statistics etc. While our method does not pinpoint the problem it prunes the corpora significantly for an action.

## 6. OUR EXPERIMENTS AND RESULTS

We have used corpora of 26000+ database queries.

The different features that we have taken are SQLtext of the query, CPU Cost, IO Cost, Memory required for intermediate results and Number of records impacted. The basis of the same are major cost components of a query cost model as discussed in section five. The execution characteristics of the queries are generally stored in system log tables for all major databases. So formulating the query to get the same is a trivial task and should be achievable for all vendors. Appropriate actual columns can be found out by refereeing the product manual and technical documentations.

We adapt a battery of methods approach as both the cost of false negative and false positive might very high. Where the cost of false positive (Actually normal but classified as outlier) is higher, we can use an intersection of the results to ensure lesser miss. Whereas when the cost of false negative (Actually an outlier but classified as normal) is high, we can use union of the results. The success of the combination will depend on the diversity of the detectors. Needless to say all the techniques we used are unsupervised as we do not have any labeled data.

### 6.1 Distance Based Approach

We picked up the five attributes as discussed in the previous section. As a first step we normalized the data points. As otherwise different attributes would have got different importance. For normalizing, we have used median and inter-quartile range, rather than mean and standard deviation, as the former two are more robust with higher breakdown point. The steps are as below

Step 1: Compute median and quartiles for each of the dimensions or features

Step 2: Normalize the feature value for all the data instances.

Step 3: Compute Euclidian distance for each of the points.

Step 4: Sort by distance in descending order

Step 5: Pick up top r %

As an improvement point we can use Mahalnabis distance to factor interdependence among the attributes. Also a precise definition of r is elusive. Following is a summary of the distribution in table 1.

Needless to say, we need to concentrate on 50+ populations. While we examined the queries we saw complete cross-products, usage of functions in where, use of derived tables, use of analytical functions, usage of not in or not exists predicate etc. Some of the queries looked inexpensive yet had high distance necessitating introspection on the data distribution and accuracy of statistics.

**Table 1. Data Distribution**

| Distance Band | No. of Observation |
|---|---|
| 1-50 | 26141 |
| 51-100 | 265 |
| 101-150 | 53 |
| 151-200 | 33 |
| 201-250 | 8 |
| 250+ | 8 |
| Grand Total | 26508 |

Two important observations here are, firstly there can be some queries as outliers because of the sheer volume of the underlying relations. Secondly the ones within the vicinity of origin cannot be conclusively said to be conforming to best practice. Rather the adverse effect of these queries is not so high so can be temporarily ignored. There are two terms known as masking and swamping with respective to outlier. Often so happens, one point gets classified as an outlier only when another outlier is removed. In this case, the former point is said to be masked by the second point. Swamping is just the other way round; a point is classified as an outlier on the presence of an outlying observation. For more details [17] can be referred. So basically even some outliers are masked. First we can remove the outliers with highest outlyingness and these masked outliers with relatively lesser outlyingness will be pronounced.

## 6.2 Clustering Based Approach

In this approach, we use the traditional K-Means. However a big challenge here is determination of K. We can refer to [18] for a solution of the same. Though objective of clustering is partitioning data set in similar groups we can identify observations as outliers by identifying clusters with minimal population. The below is the frequency distribution of the clusters. (By Default setting we have taken K as 10). Figure 3 is a bar chart with the clusters and their corresponding count.

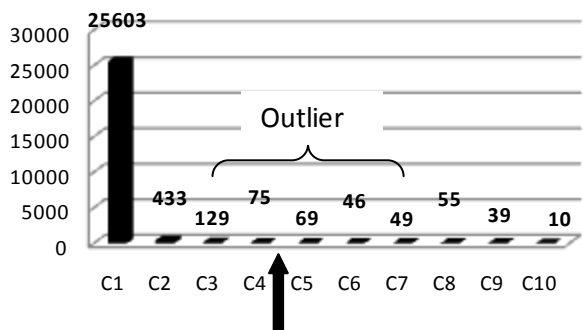

**Figure 3: Cluster wise Distribution**

The arrow in figure 3 is more of a slider which will be requirement driven. Clustering is one of very popular methods for data mining tasks. Some of the popular algorithms are K-Means, Nearest Neighbor, PAM etc. For larger data sets, sampling based approaches like CLARA or CLARANS are used. For a detailed overview we can refer to [19].

## 6.3 Average Distance Based approach

The basic philosophy is if an instance is on average at a distance higher than the other points then it is an outlier.

Step 1: Start with normalizing the values of all the features.

Step 2: Find distance between all points. (The problem of this approach is time complexity.)

Step 3: Calculate average distance with neighbors.

Step 4: Pick up the top n%, sorting by the average distance.

Table 2 shows the distribution as per distance band.

**Table 2: Data distribution as per average distance**

| Average Distance Band | No. of Observation |
|---|---|
| 0-50 | 26309 |
| 51-100 | 144 |
| 101-150 | 40 |
| 151-200 | 9 |
| 200+ | 6 |
| 0-50 | 26309 |
| Grand Total | 26508 |

## 6.4 Density Based Approach

There can be a dense population away from the center of the data. Distance based outliers will end up classifying them as outliers. A density based approach attempt to overcome the same. A density based approach looks at regions as a dense or a sparse region. The population at a sparse region is identified as outliers. We use local outlier factor (LOF) for the same, details of the same can be found in [20]. It observes the global view taken is meaningful and adequate under certain condition but not satisfactory for general condition where clusters of different density exist. So this is optimized for not only identifying global outliers but local outliers as well. We give a simplified account of the algorithm by taking K as 10.

Step 1: For each object find a distance D, such that there are at least 10 neighbors within that distance.

Step 2: Find the neighbors of each point such that their distance with the point is less than or equal to D. Cardinality of such a set can be more 10. As example from a point O, there can be 8 points within a distance of 5 and then 3 points at distance 6. In this case D is 6 and the cardinality of the set is 11, rather than 10.

Step 3: Local reachability density for each point is calculated which is nothing but the inverse of average reachability distance with its neighbors. An average reachability distance is a distance metric which is calculated using the following basis

If it is less than d, then d. Else the actual distance

Step 4: A Local Outlier Factor (LOF) for each point is determined, which is average of the ratio between the neighbors' reachability index with its own.

We broke this in multiple tables and analyzed the results. The K taken in this case was 10. Interestingly query Id 26507, which we elaborate in further details in the result section was no more detected as an outlier as it has many neighbors with similar local reachability density. Hence 'K' can play a big role in this algorithm. As an improvement we can think of normalizing the local reachability density before calculating the local outlier factor. Because of the not so encouraging results we do not consider this method for further combination.

## 6.5 Combining the Approaches

[19] Describes an ensemble of outliers as different outliers are specialized for different kind of data. Various kinds of combination techniques can be employed like weighted sum, majority voting or weighted majority voting. A challenge is combining level based outliers with score based outliers. The

paper [19] shows a combination gives better result on real datasets. We propose a very simple combination framework of taking either the union or intersection, depending on the availability of IT Budget at this point of time. Needless to say a union would be more robust as far as false negatives are concerned. At the same time it will take more effort which translates to a higher dollar value. Our approach suffers from lack of calibration and any other combining function can be used, however that will be at the cost of additional computing power.

We start with a simple setting of taking first 1% as outlier, as both method one and method three produces distances. For the third one we start with the lowest populated cluster and go on taking the next cluster until the sum is more than 1%.

Step 1: Sort the cluster by population ascending

Step2: Sum = sum +Population (Clust i)

Step 3: Check if Sum > n% * Total Population

Step 4: If yes continue from Step 2 with the next cluster.

Using the same we pick-up cluster 6 to 10. The combined population is 268. We can further refine it to make it exact n% by ranking populations in the last included cluster by its distance from the cluster center. Higher the distance from the center, higher is the rank.

When we combine & analyze, the combination gives a total of 278 observations, as shown in Table 3.

**Table 3: Combination of methods**

| No. of Methods | No. of Observation |
|---|---|
| 1 | 15 |
| 2 | 15 |
| 3 | 248 |
| Grand Total | 278 |

So if we take the union method we would be dealing with all 278, where as if we take an intersection we would go with 248. There can't be confidence attached with the detectors as firstly this is an unsupervised method and secondly verification of its outlyingness is time consuming & dependent on technical experts.

## 6.6 Analyzing the Results

We inspect and sight the cases that we have seen. The exact relation names are confidential; hence we will be using fictions aliases. We start examining the query with highest distance from origin.

```
SELECT SRC.A , TGT.A FROM ( SELECT A FROM T1
WHERE Batch = 1678 ) SRC

INNER JOIN ( SELECT A FROM T2 WHERE Batch = 1668
) TGT

ON SRC.A <> TGT.B ;
```

The above query is surely incorrectly formulated. Looking at the same we can understand the objective is to find out values of Column A which exist only in one table. However the above query will result in almost a cross join. Hence the cost of the same is very high. Next five observations are instances of the same query. The same query is topping in the average distance. The same query is also appearing in the cluster with the lowest population that is cluster 10.

We examine the next one and following is the query.

```
SELECT

COALESCE ( ( ADD_MONTHS ( a.Some_date , ( C.Vrsn -
Vrsn ) * - 12 )

( FORMAT 'MMDDYYYY' ) ( CHAR ( 8 ) ) ) , ' ' )

FROM T1 a , T2 b , T3 c

WHERE SUBSTR ( TRIM ( A.Field ) , 1 , 1 )   IN
'4' ;
```

At a first look we might feel, this is a case where the query is not hitting the index and going for a full table scan because a function is used on the column. However a careful inspection reveals three tables are being used without any join condition. As the earlier case the same query is appearing in cluster 10 and in top ranks of average distance.

The detectors that we use are more of similar nature hence the intersection is high, the diverse the detectors the more benefit we achieve as there is a cost involved in using more than one detectors and then combining them.

A very important understanding here is though LOF has a much sounder foundation it did not seem to give correct result, which reinforces the view of choosing detection methods which are suited for the particular application domain. As example there can be a query which is very infrequent and rare in its kind and can be identified as an outlier by LOF. However this are not of the interest as far as resource consumption and query optimization is concerned. As mentioned in the motivation profiling of queries are also going to be very important and here approaches like LOF will be beneficial, however as here our focus is more to find costlier queries, this method did not produce good results.

## 7. CONCLUSION

In this paper we have taken a very different approach to query optimization. While standard techniques are applied while formulating the query, in reality the cost of queries changes significantly because of various reasons. We propose a framework here for identifying queries with highest distance from the group and then taking corrective action. We have proposed three methods for the same and suggested very simple way of combining them however even individually all three methods are effective.

As a next step, we plan to study the temporal behavior of the queries as well. We would like to find best performing queries for an imitable best practice. We would also work on making our combining framework more robust and formal. We would also like to use it for a better insight in the executed queries as understanding of workloads is getting prominence [1]. This will enable an optimal planning for capacity as well as scheduling.

In this paper, we take a unique approach towards query tuning. Though this is a much matured space, our approach is very different from the existing ones. The focus of the paper is also completely aligned with the major trends as reckoned by Gartner [1].This offers immediate benefits in terms of reduction of information latency as well as an optimal use of hardware and software.

# 8. REFERENCES

[1] Donald Feinberg, Mark A. Beyer , Gartner RAS Core Research Note G00209623, 28 January 2011

[2] J. Hodge (vicky@cs.york.ac.uk)∗ and Jim Austin , "A Survey of Outlier Detection Methodologies" Victoria Artificial Intelligence Review, 2004

[3] V Chandola, A Banerjee, B Kuman, Anomaly Detection: A Survey, ACM Computing Survey 2009

[4] P. Filzmoser, R. Maronna, and M. Werner, "Outlier identification in high dimensions", Computational Statistics and Data Analysis , Volume 52, Issue 3, 1 January 2008, Pages 1694-1711

[5] S. Subramaniam et. al, "Online outlier detection in sensor data using non-parametric models", VLDB '06 Proceedings of the 32nd international conference on Very large data bases

[6] Clifton Phua, Vincent Lee, Kate Smith, Ross Gayler, "A Comprehensive Survey of Data Mining-based Fraud Detection Research", arXiv:1009.6119v1

[7] Wenke Lee and Salvatore J. Stolfo ,"Learning Patterns from Unix Process Execution Traces for Intrusion Detection" AAAI Workshop on AI Approaches to Fraud Detection, 1997

[8] Surajit Chaudhuri , "An overview of query optimization in relational systems" , PODS '98 Proceedings of the seventeenth ACM SIGACT-SIGMOD-SIGART symposium on Principles of database system

[9] Matthias Jarke and Jurgen Koch, "Query Optimization in Database Systems" , ACM Computing Surveys (CSUR) Surveys Homepage archive Volume 16 Issue 2, June 1984

[10] Nicolas Bruno, Surajit Chaudhuri and Ravi Ramamurthy, "Power Hints for Query Optimization", Data Engineering, 2009. ICDE '09. IEEE 25th International Conference

[11] Yuqing Wu, Jignesh M. Patel and H. V. Jagadish, "Structural Join Order Selection for XML Query Optimization", 19th International Conference on Data Engineering (ICDE'03)

[12] Samiran Ghosh, Saptarsi Goswami, Amlan Chakrabarti , "Outlier detection from ETL Execution trace", 2011 International Conference on Network and Computer Science (ICNCS 2011)

[13] Barnett, V. and Lewis, T.: 1994, "Outliers in Statistical Data." John Wiley and Sons.,3 edition.

[14] Hawkins D, "Identification of Outliers", Chapman and Hall, 1980

[15] Kimbal, Ralph and Caserata, Joe,"The Datawarehouses ETL Tool Kit." 1. s.l. : Wiley. p. 528. 978-0764567575.

[16] Lucantonio Ghionna et. al ,"Outlier detection techniques for process mining applications," ISMIS'08: Proceedings of the 17th international conference on Foundations of intelligent systems"

[17] Irad Ben-Gal, "Outlier Detection", Data Mining and Knowledge Discovery Handbook, 2010

[18] Ujjwal Das Gupta, Vinay Menon, Uday Babbar., "Detecting the number of clusters during Expectation-Maximization clustering using Information Criterion", 2010 Second International Conference on Machine Learning and Computing

[19] Rui Xu; Wunsch, D., II, "Survey of clustering algorithms", IEEE Transactions on Neural Networks

[20] Markus M. Breunig et. al , "LOF: identifying density-based local outliers", SIGMOD '00 Proceedings of the 2000 ACM SIGMOD international conference on Management of data

[21] Hoang Vu Nguyen, Hock Hee Ang and Vivekanand Gopalkrishnan, "Mining Outliers with Ensemble of Heterogeneous Detectors on Random Subspaces", Database Systems for Advanced Applications, 2010.